\documentclass[pre,twocolumn]{revtex4}
\usepackage{amssymb}
\usepackage{color}

\usepackage{amsthm}
\usepackage{graphicx}
\usepackage{dcolumn}
\usepackage{bm}
\usepackage{subfigure}
\usepackage{amssymb}
\usepackage{verbatim}
\usepackage{amscd}
\usepackage{amssymb}
\usepackage{amsmath, amsfonts}
\usepackage{setspace}
\usepackage[]{graphicx}        
\usepackage{amsthm}
\usepackage{natbib}
\usepackage{enumerate}

\newcommand{\ba}[1]{\begin{align*} #1 \end{align*}}

\theoremstyle{plain}

\theoremstyle{plain}

\theoremstyle{plain}
 
\theoremstyle{plain}

\theoremstyle{remark}

\theoremstyle{conjecture}

\theoremstyle{observation}

\theoremstyle{definition}

\theoremstyle{corollary}

\theoremstyle{definition}

\theoremstyle{definition}

\theoremstyle{assumption}

\theoremstyle{definition}

\theoremstyle{problem}

\theoremstyle{fact}

\begin{document}

\title{Precise estimation of critical exponents from real-space renormalization group analysis}
\author{Aleksander Kubica}
\author{Beni Yoshida}
\affiliation{Institute for Quantum Information and Matter, California Institute of Technology, Pasadena, California 91125, USA
}

\date{\today}
\begin{abstract}
We develop a novel real-space renormalization group (RG) scheme which accurately estimates correlation length exponent $\nu$ near criticality of higher-dimensional quantum Ising and Potts models in a transverse field. Our method is remarkably simple (often analytical), grouping only a few spins into a block spin so that renormalized Hamiltonian has a closed form. A previous difficulty of spatial anisotropy and unwanted terms is avoided by incorporating rotational invariance and internal $\mathbb{Z}_q$ symmetries of the Hamiltonian. By applying this scheme to the (2+1)-dim Ising model on a triangular lattice and solving an analytical RG equation, we obtain $\nu\approx 0.6300$. This value is within statistical errors of the current best Monte-Carlo result, $25$th-order high-temperature series expansions, $\phi^4$-theory estimation which considers up to seven-loop corrections and experiments performed in low-Earth orbits. We also apply the scheme to higher-dimensional Potts models for which ordinary Monte-Carlo methods are not effective due to strong hysteresis and suppression of quantum fluctuation in a weak first-order phase transition.
\end{abstract}
\maketitle

\section{Introduction}

Quantum phase transitions occur when ground state properties of interacting many-body systems dramatically change under tiny changes of parameters in a parent Hamiltonian at zero temperature~\cite{Sachdev_Text}. A remarkable prediction is that the universality class of phase transitions can be completely characterized by a set of critical exponents which encode how physical observables change across the transition point. However, finding critical exponents is a problem of tremendous analytical and computational difficulty as it involves an exponentially large Hilbert space due to diverging correlation length. 

\begin{figure}[htb!]
\centering
\includegraphics[width=0.75\linewidth]{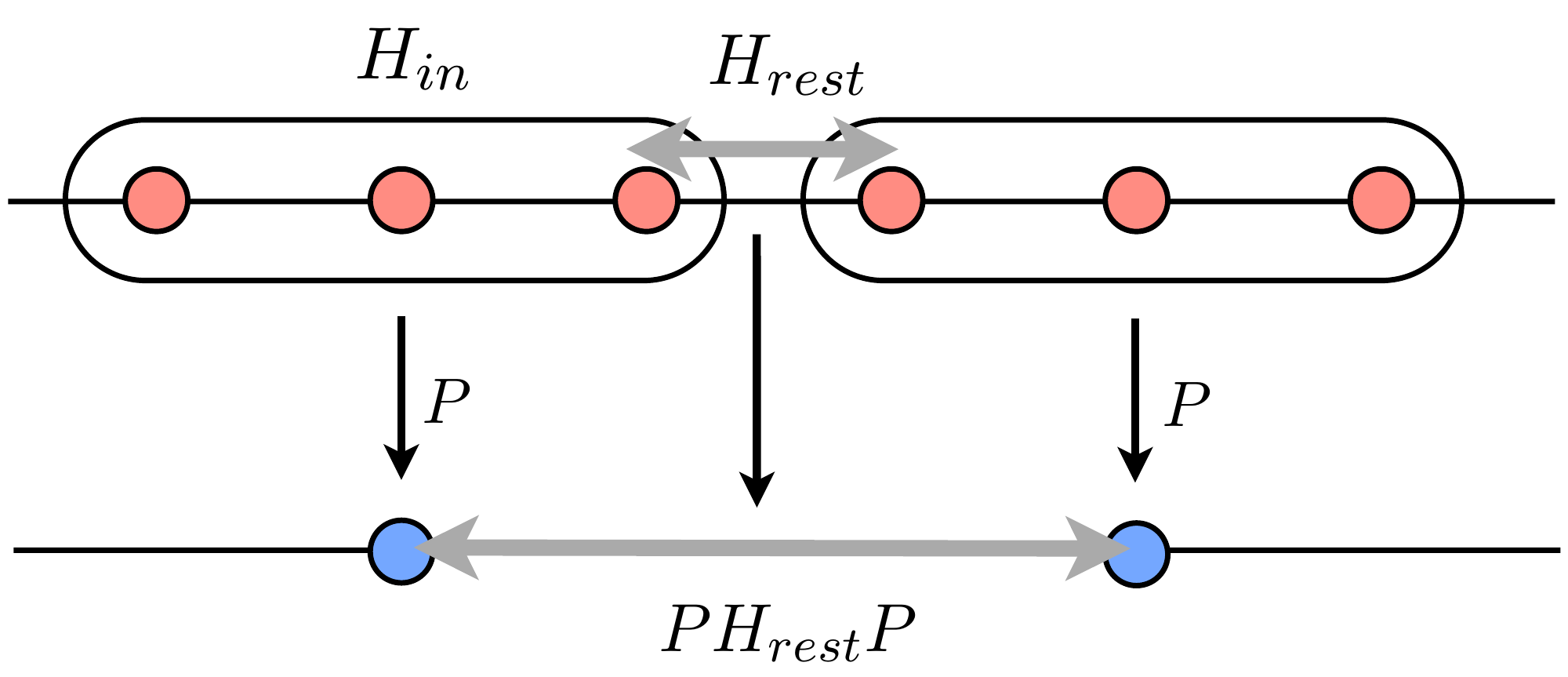}
\caption{(Color online) Real-space RG transformation. A block of three spins is renormalized into a single spin in this example.  
} 
\label{fig_RG}
\end{figure}

A traditional approach to study quantum critical phenomena is via real-space renormalization group (RG) transformation~\cite{KADANOFF66}. To be specific, consider a spin-$1/2$ model governed by some Hamiltonian $H$ with local interactions (Fig.~\ref{fig_RG}). We divide the lattice into blocks of spins and split the Hamiltonian into two parts:
\ba{
H = H_{rest} + \sum_{I \in blocks} H^{I}_{in} 
}
where $H^{I}_{in}$ consists of all the terms acting only within the block $I$ and $H_{rest}$ consists of all the other terms. Within each block $I$, we compute the lowest two ground states $|\psi^{I}_{0}\rangle$ and $|\psi^{I}_{1}\rangle$ of $H^{I}_{in}$ and treat them as basis states forming a renormalized spin. We then obtain a renormalized Hamiltonian
\ba{
H'= PH_{rest}P + \sum_{I \in blocks}\left( e_{0}|\psi^{I}_{0}\rangle\langle  \psi^{I}_{0}|+ e_{1}|\psi^{I}_{1}\rangle\langle  \psi^{I}_{1}|\right)
}
where $P$ is a projector onto a space spanned by $|\psi^{I}_{0}\rangle$ and $|\psi^{I}_{1}\rangle$, and $e_{0}$ and $e_{1}$ are energies of $|\psi^{I}_{0}\rangle$ and $|\psi^{I}_{1}\rangle$. Due to the scale invariance emerging at quantum criticality, we require that the renormalized Hamiltonian $H'$ and the original Hamiltonian $H$ have a similar form. By investigating how coupling strengths change from $H$ to $H'$ and solving for fixed points of RG equations, we may analyze a quantum phase transition.

However, this real-space RG approach generally does not give precise estimate of critical exponents. For instance, when applied to the quantum Ising model in one dimension, estimations deviate from the correct values by around 20 percent even when more than ten spins are grouped into a block spin, and estimations do not converge to exact values even with the increasing block size~\cite{Jullien78}. The situation becomes even worse in higher dimensions where the renormalized Hamiltonian $H'$ often consists of unwanted terms that were not present in the original Hamiltonian $H$~\cite{Penson79}. These difficulties have severely limited the usage of real-space RG method in estimating critical exponents. 

Recently, Miyazaki \textit{et al} proposed that, instead of including all the terms in $H_{in}$, other choices of $H_{in}$ may give far better estimations of critical exponents~\cite{Miyazaki11}. In particular, they pointed out that, in the quantum Ising model, there exist choices of $H_{in}$ whose estimation accuracy of correlation length exponent $\nu$, defined as
\begin{align}
\xi \sim |\lambda - \lambda_{0}|^{-\nu},
\end{align}
is $3-5$ percent, where $\lambda$ is some order parameter. Unfortunately, estimation of the magnetic exponent $\beta$, defined as $m \sim |\lambda - \lambda_{0}|^{\beta}$, is not precise. Yet, since there are only two independent critical exponents under scaling relations, this result opens a possibility of developing a novel real-space RG scheme which is \emph{tailor-made} to precisely (or even analytically) compute correlation length exponents. 

In this paper, we develop a novel real-space RG scheme which is able to predict values of $\nu$ in higher-dimensional quantum Ising and Potts model. Our scheme, when applied to the two-dimensional quantum Ising model, gives an estimate consistent with the current best Monte-Carlo simulations which typically require $10-100$ CPU years~\cite{Hasenbusch10}, estimates from $25$th-order high-temperature series expansions~\cite{Campostrini02}, $\phi^4$-theory with seven loop corrections with highly intricate resummation schemes~\cite{Kleinert99} and experiments performed in low-Earth orbits where gravity-induced variations are small~\cite{Barmatz07}. For details of previous estimates, we refer the reader to~\cite{Hasenbusch10, Pelissetto02} and references therein. We also apply this method to higher-dimensional quantum Potts model where Monte-Carlo method does not work efficiently due to weak first-order phase transition~\cite{Blote79, Binder87}. 

\section{One-dimensional real-space RG}

We begin with real-space RG analysis of one-dimensional quantum Ising model in a transverse field:
\begin{align}
H_{Ising} = - J\sum_{j}Z_{j}Z_{j+1} - h\sum_{j}X_{j}
\end{align}
where $J,h>0$. A traditional approach is to include all the terms inside a block in the inner Hamiltonian $H_{in}$. When a block consists of two spins, one has 
\begin{align}
H_{in}=-JZ_{1}Z_{2}-hX_{1}-hX_{2}
\end{align}
where $Z_{j}$ and $X_{j}$ act on $j$th spin in a block ($j=1,2$). It is well known that the above approach does not give precise estimate of critical exponents, and is only able to predict the presence of quantum phase transition. 

Instead, Miyazaki \textit{et al}~\cite{Miyazaki11}, based on a pioneering result by  Fernandez-Pacheco~\cite{Fernandez-Pacheco79}, has suggested that a different choice of $H_{in}$ may significantly improve the precision: by removing one of the on-site $X$ terms, namely 
\begin{align}
H_{in}=-JZ_{1}Z_{2}-hX_{1}.  \label{eq:1dim}
\end{align}
After the renormalization step, the ferromagnetic coupling and on-site transverse field rescale as follows:
\begin{align}
J' = \frac{J^2}{\sqrt{J^2+h^2}} \qquad h' = \frac{h^2}{\sqrt{J^2+h^2}}\label{eq:Ising}
\end{align} 
by projecting $H_{rest}$ into low energy subspace of $H_{in}$. This allows us to compute the correlation length exponent $\nu$ as follows
\begin{align}
\nu^{-1}=\log_2 \left.\frac{d(h'/J')}{d(h/J)}\right|_{h/J=h'/J'}= \log_2 \left.\frac{dk'}{dk}\right|_{k=k'} \label{eq:RG}
\end{align}
where
\begin{align}
k\equiv\frac{J}{h} \qquad k'\equiv\frac{J'}{h'}.
\end{align}
The base of the logarithm is two since we rescaled the system by the factor of two. Surprisingly, the above RG equation, applied to Eq.~(\ref{eq:Ising}), gives $\nu=1$, which is \emph{precisely equal} to the exact solution. 

However, at this moment, it is unclear why the above RG scheme works well. After all, this may be just a coincidence! Are there any conditions on choices of $H_{in}$ under which the RG scheme works well? Is the method applicable to other models of quantum phase transition? What is the reason behind the success of the above RG scheme? Below we address these questions. 

Via throughout investigations of various $H_{in}$ in one dimensions, we found that the most important condition in choosing $H_{in}$ is to guarantee that it has degenerate ground states and respects $\mathbb{Z}_{2}$ symmetry. In the RG scheme by Miyazaki \textit{et al}, one may observe the degeneracy in $H_{in}$ by considering symmetry operators $\bar{X}$ and $\bar{Z}$:
\begin{align}
H_{in} = - J Z_{1}Z_{2} - hX_{1}, \quad \bar{X} \equiv X_{1}X_{2} \quad \bar{Z}\equiv Z_{2} \label{eq:symmetry}
\end{align}
which commute with $H_{in}$, and form an algebra of Pauli operators on degenerate ground states:
\ba{
[H_{in},\bar{X}]=[H_{in},\bar{Z}]=0,\qquad \{ \bar{X},\bar{Z}\}=0
}
where two ground states $|\tilde{0}\rangle$ and $|\tilde{1}\rangle$ obey $\bar{Z} |\tilde{0}\rangle = + |\tilde{0}\rangle$, $\bar{Z} |\tilde{1}\rangle = -|\tilde{1}\rangle$, $\bar{X} |\tilde{0}\rangle = |\tilde{1}\rangle$, $\bar{X} |\tilde{1}\rangle = |\tilde{0}\rangle$, and $[A,B]$ represents a commutation and $\{A,B\}$ represents an anti-commutation between operators $A$ and $B$.

\begin{figure}[htb!]
\centering
\includegraphics[width=0.80\linewidth]{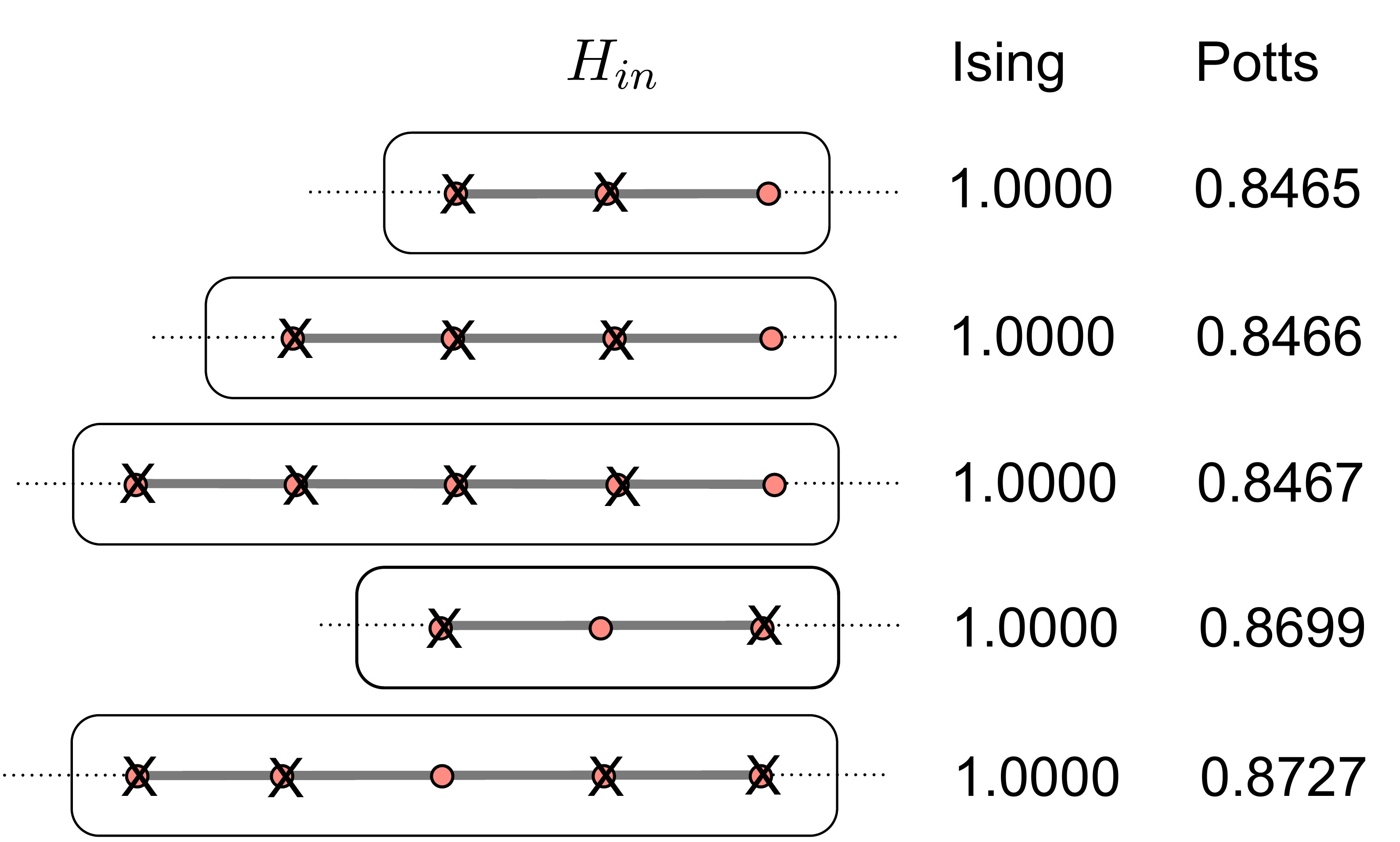}
\caption{(Color online) Estimation of correlation length exponent $\nu$ from real-space RG with degenerate $H_{in}$ for the quantum Ising and $3$-state Potts models. The figure shows the choices of $H_{in}$ where crosses represent $X$ terms and solid lines represent $ZZ$ terms. In these examples, all the ferromagnetic terms inside a block are included in $H_{in}$.
} 
\label{fig_1Dlist}
\end{figure}

To verify the importance of the degeneracy in $H_{in}$, we employed various choices of degenerate $H_{in}$ as shown in Fig.~\ref{fig_1Dlist} where crosses represent on-site $X$ terms and solid lines represent ferromagnetic $ZZ$ terms. For all the choices depicted in Fig.~\ref{fig_1Dlist}, we found $|\nu -1| <10^{-7} $ by numerically evaluating RG equations. We expect that $\nu=1$ precisely on these RG equations though we do not have analytical proof on this observation. Thus, $\mathbb{Z}_{2}$ symmetry of the inner Hamiltonian $H_{in}$ seems to be the crucial condition for successful RG schemes. 

A naturally arising question concerns why degenerate $H_{in}$ is favorable in computing critical exponents. An important observation is that, under degenerate $H_{in}$, the renormalized Hamiltonian $H'$ always has a closed form with terms proportional to either ferromagnetic $ZZ$ or on-site magnetic field $X$, up to some additive constant corrections. This statement can be rigorously proven for the quantum Ising and Potts model on an arbitrary lattice by using algebraic properties of Pauli operators. For simplicity of discussion, we demonstrate that the RG scheme, considered by Miyazaki \textit{et al}, generates no extra term. Let $P$ be a projector onto two degenerate ground states of $H_{in}=-JZ_{1}Z_{2}-hX_{1}$. The action of Pauli operator $Z_{1}$ inside the ground state space is
\ba{
&\langle \tilde{0}| Z_{1} | \tilde{0}\rangle =\langle \tilde{1}| \bar{X}Z_{1} \bar{X} | \tilde{1}\rangle = -\langle \tilde{1}| Z_{1} | \tilde{1}\rangle \\
&\langle \tilde{0}| Z_{1} | \tilde{1}\rangle = \frac{1}{4}\langle \tilde{0}| (1+\bar{Z})Z_{1}(1-\bar{Z}) | \tilde{1}\rangle = 0
}
so, $PZ_{1}P \sim \bar{Z}$. Similar reasoning leads to $PZ_{2}P \sim \bar{Z}$ and $PX_{2}P \sim \bar{X}$. Therefore, the renormalized Hamiltonian has a closed form. This analysis can be straightforwardly generalized to arbitrary $H_{in}$ with $\mathbb{Z}_{2}$ symmetry, leading to a closed form of $H'$.

Next, in order to check the universality of the method, we study the quantum $3$-state Potts model:
\begin{align}
H_{Potts} = - J\sum_{j} Z_{j}^{\dagger} Z_{j+1}- h\sum_{j}X_{j} + \mbox{h.c.}
\end{align}
where $Z_{j}$ and $X_{j}$ are generalized Pauli operators defined for three-dimensional spins. By applying real-space RG with
\begin{align}
H_{in} = - J (Z_{1}^{\dagger}Z_{2}+Z_{1}Z_{2}^{\dagger}) - h(X_{1}+X_{1}^{\dagger})
\end{align}
we obtain
\begin{align}
k\rightarrow k'=k\frac{\sqrt{9k^2-6k+9}+9k-3}{\sqrt{9k^2-6k+9}-3k+9}
\end{align}
where $k\equiv J/h$ and $k' \equiv J'/h'$. From Eq~(\ref{eq:RG}) we calculate
\begin{equation}
\nu =1/\log_2 (4-\sqrt{3})\approx 0.8464
\end{equation}
which is close to the exact value $\nu=5/6 \approx 0.8333$ obtained from conformal field theory. 

Note that $H_{in}$ has three degenerate ground states and has $\mathbb{Z}_{3}$ symmetry where $\bar{X}=X_{1}X_{2}$ and $\bar{Z}=Z_{2}$ are symmetry operators. In order to further verify the importance of degenerate $H_{in}$, we also analyze the quantum Potts model via other choices of degenerate $H_{in}$ with $\mathbb{Z}_{3}$ symmetries. The results are listed in Fig.~\ref{fig_1Dlist} which agree well with the exact value. Yet, there are a few important distinctions. First, increasing the block size does not improve the estimation in general. Second, removing the $X$ term at the boundary gives better estimates. We will return to these two observations by analyzing two-dimensional real-space RG schemes.

\section{Two-dimensional real-space RG}

In one dimension, we have found that degenerate $H_{in}$ generally gives a precise estimate of $\nu$ since the renormalized Hamiltonian remains closed for both quantum Ising and Potts models. Here, we proceed to develop higher-dimensional real-space RG schemes for the quantum Ising and Potts models. 

\begin{figure}[htb!]
\centering
\includegraphics[width=0.85\linewidth]{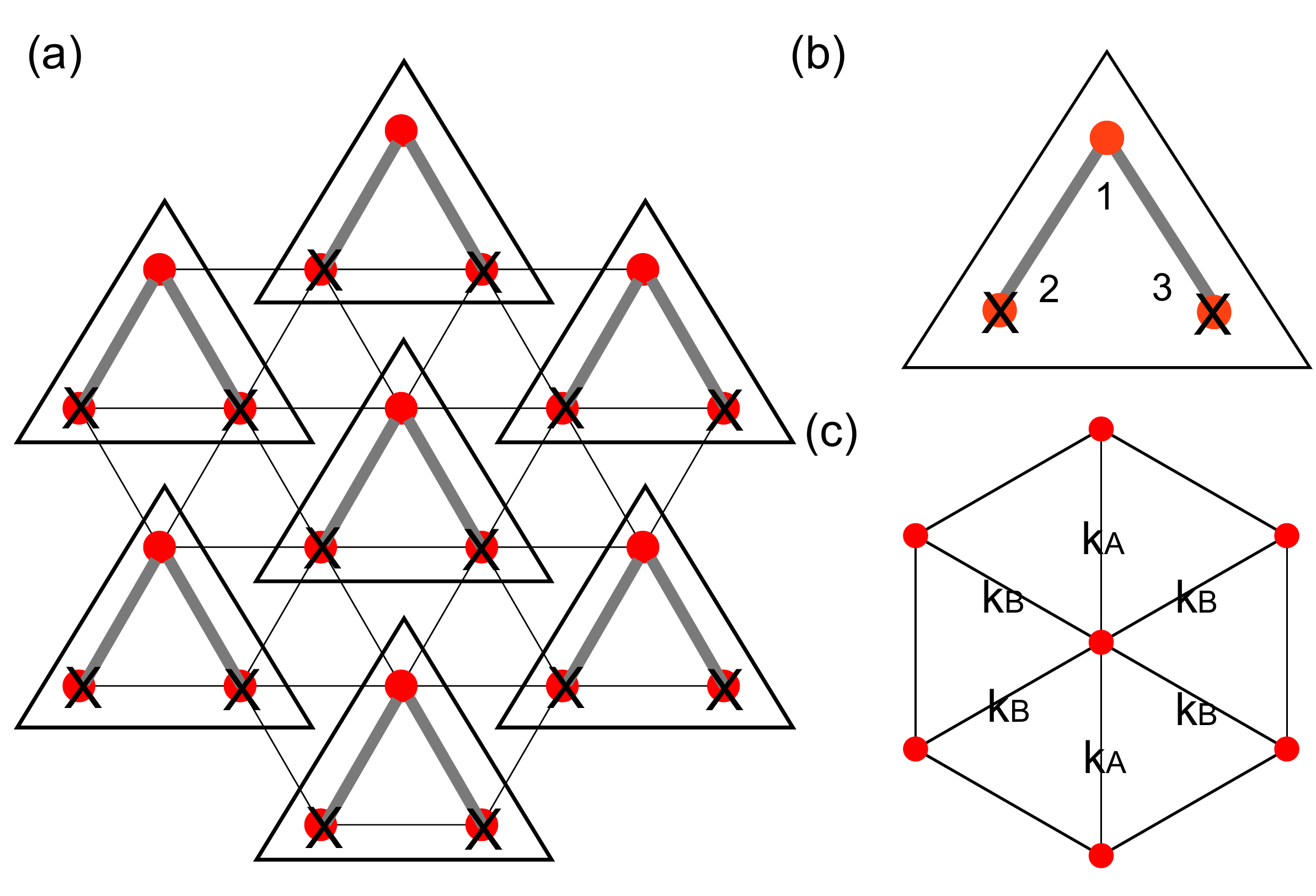}
\caption{(Color online) Renormalization of the quantum Ising model on a triangular lattice. (a) The covering. (b) The inner Hamiltonian $H_{in}$. (c) Renormalized coupling strengths.
} 
\label{fig_triangular}
\end{figure}

An immediate challenge in two dimensions is that the renormalized Hamiltonian becomes anisotropic since the inner Hamiltonian $H_{in}$ is not isotropic in general. To be specific, let us consider two-dimensional quantum Ising model defined on a triangular lattice. We think of covering the entire system by blocks of three spins, and choosing $H_{in}$ as depicted in Fig.~\ref{fig_triangular}(a)(b):
\begin{align}
H_{in} = -J(Z_{1}Z_{2} +Z_{1}Z_{3}) - h (X_{2}+X_{3}).
\end{align}
Note that $H_{in}$ has $\mathbb{Z}_{2}$ symmetry with
\begin{align}
\bar{X}\equiv X_{1}X_{2}X_{3} \qquad \bar{Z} \equiv Z_{1}.
\end{align}
By projecting $H_{rest}$ onto the ground space of $H_{in}$, we obtain coupling terms which bridge neighboring blocks (renormalized spins). Note there are six bonds outgoing from a single renormalized spin and two types of coupling strengths $k'_A$ and $k'_B$
\begin{align}
k'_{A}& = 2k^2 \sqrt{k^2+1}\\
k'_{B}& = k^2 (k+\sqrt{k^2+1})
\end{align}
as depicted in Fig.~\ref{fig_triangular}(b). Therefore, RG equations become highly anisotropic.  

\begin{figure*}[htb!]
\centering
\includegraphics[width=0.75\linewidth]{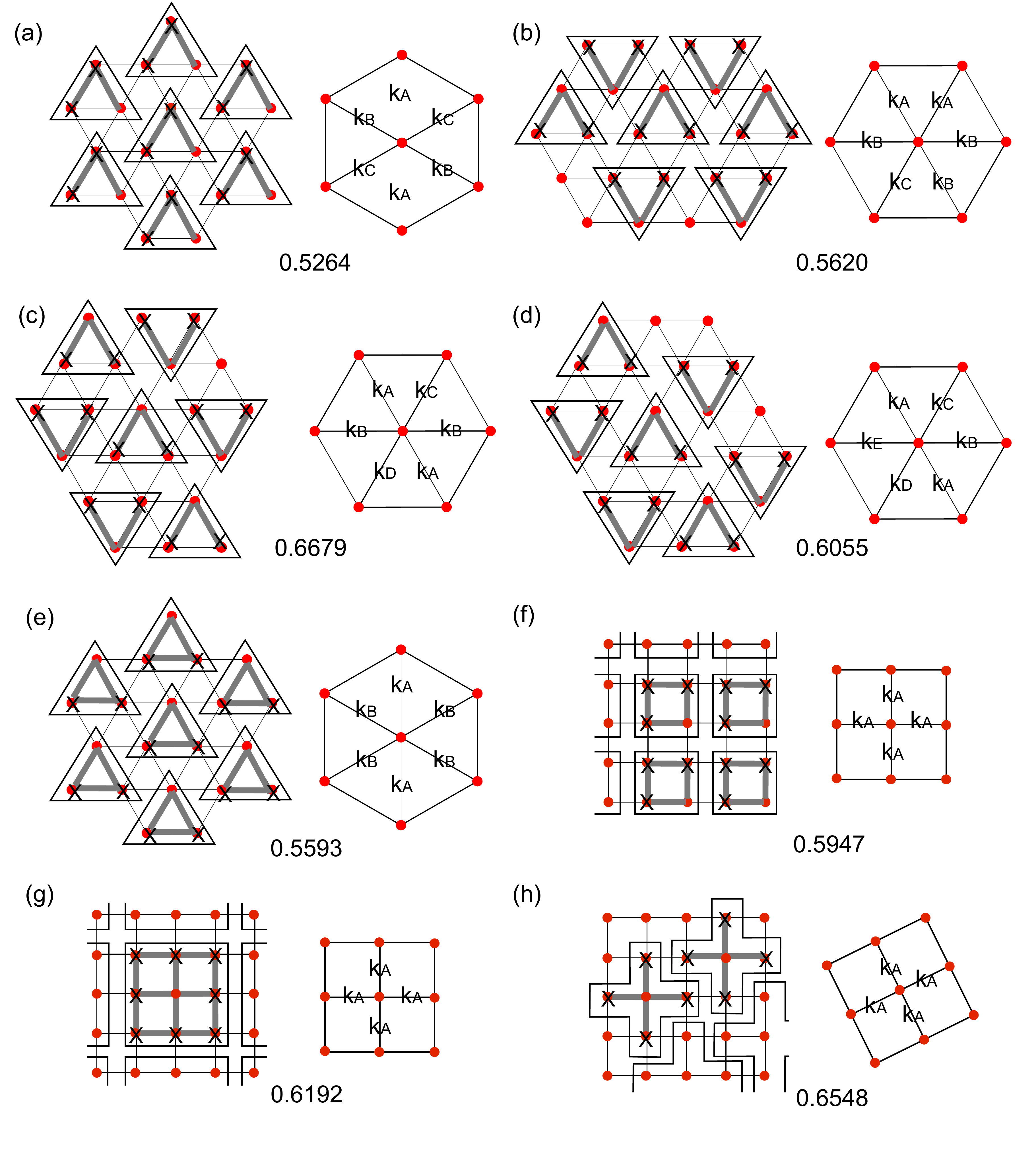}
\caption{(Color online) Estimations of correlation length exponent $\nu$ from various choices of $H_{in}$.
} 
\label{fig_misc}
\end{figure*}

In the case of anisotropic coupling strengths $(k'_{A},k'_{B})$, the standard approach \cite{Miyazaki11} is to linearize the renormalization group transformation, find its eigenvalues and keep only the biggest one. This procedure is not fully justified since by discarding the other eigenvalue we lose some information how the system rescales. In what follows, we propose that the renormalized coupling strength should be a geometric mean of all coupling strengths. For the considered example of a triangular lattice we have
\begin{align}
k' \equiv (k'_{A})^{2/6}(k'_{B})^{4/6}.
\end{align}
We then obtain an RG equation
\begin{align}
k\rightarrow k'  = k^2 (2\sqrt{k^2+1})^{1/3} (k+\sqrt{k^2+1})^{2/3}
\end{align}
which has a non-trivial fixed point. We obtain the critical exponent to be
\begin{align}
\nu \approx 0.6300
\end{align}
which is within a statistical error of the current best Monte-Carlo estimation, the $\phi^4$-theory estimation, the high-temperature series expansion and experiments. Therefore, real-space RG is able to predict $\nu$ with surprisingly good precision in two dimensions, too. We shall call this real-space RG scheme the \empty{canonical method}. 

In order to understand why the canonical method works well, we have investigated a number of other choices of inner Hamiltonian $H_{in}$ for various lattices. We find a few important conditions under which real-space RG works well. One condition is to choose $H_{in}$ so that $H'$ is less anisotropic. We show results of real-space RG with anisotropic coupling strengths in Fig.~\ref{fig_misc}(a)(b)(c)(d). In Fig.~\ref{fig_misc}(a), three types of coupling strengths are generated after renormalization. In Fig.~\ref{fig_misc}(b)(c)(d), the number of bonds bridging the triangular blocks vary in different directions, and the renormalized coupling strengths are highly anisotropic. As expected, estimates from these RG schemes are not close to an actual value. Another important condition is to avoid making ``loops'' of ferromagnetic interactions inside a block as seen in Fig.~\ref{fig_misc}(e)(f)(g). In Fig.~\ref{fig_misc}(e), all three ferromagnetic terms are included in $H_{in}$ which form a loop. Similarly, in Fig.~\ref{fig_misc}(f), a loop is formed inside a square block. The inner Hamiltonian $H_{in}$ can be made fully isotropic by removing the on-site term $X$ only at the center, as in Fig.~\ref{fig_misc}(g), but ferromagnetic terms form multiple loops. Since we are interested in how correlations grow over the whole lattice, it seems legitimate to avoid loops which would suppress propagations of correlations. Estimates from RG schemes with loops are not close to an actual value. Also, a choice in Fig.~\ref{fig_misc}(g) seems not optimal due to observations obtained from analysis of one-dimensional Potts model since it groups a large number of spins and removes the $X$ term at the center instead of at the boundary. Finally, Fig.~\ref{fig_misc}(h) covers the lattice in a skewed way while the renormalized coupling strengths remain isotropic. Again, the estimate is not close to an actual value. Therefore, it seems to us that the canonical scheme is the only sensible choice one could consider for precise estimation of $\nu$. 

Having developed two-dimensional real-space RG scheme, let us consider an application to the quantum Potts model. In one dimension, we have seen that real-space RG is able to predict $\nu$ for the Potts model with good accuracy. In two spatial dimensions, the quantum Potts model is known to undergo first-order phase transition~\cite{Wu82}, and is not at a true quantum criticality as the scaling relations are not satisfied. Yet, the transition has some reminiscent of quantum fluctuations where correlation length diverges at the critical point and critical exponents can be defined. However, for these \emph{weak} first-order phase transitions, finding critical exponents is a problem of tremendous computational difficulty with no established method since standard Monte-Carlo method does not work effectively due to strong hysteresis and suppression of quantum fluctuations. 

Here, we obtain an estimate of correlation length exponent of the two-dimensional quantum Potts model via the canonical method. A similar analysis, involving the geometric mean, leads to
\begin{align}
\nu\approx 0.5473.
\end{align}
The current known estimate from Monte-Carlo simulation is $\nu\approx 0.40\pm0.13$~\cite{Herrmann79} which has a large uncertainty due to the difficulties mentioned above. It may be interesting to check our prediction with a large-scale numerical simulations.

\section{Hybridized real-space RG}

\begin{figure}[htb!]
\centering
\includegraphics[width=0.75\linewidth]{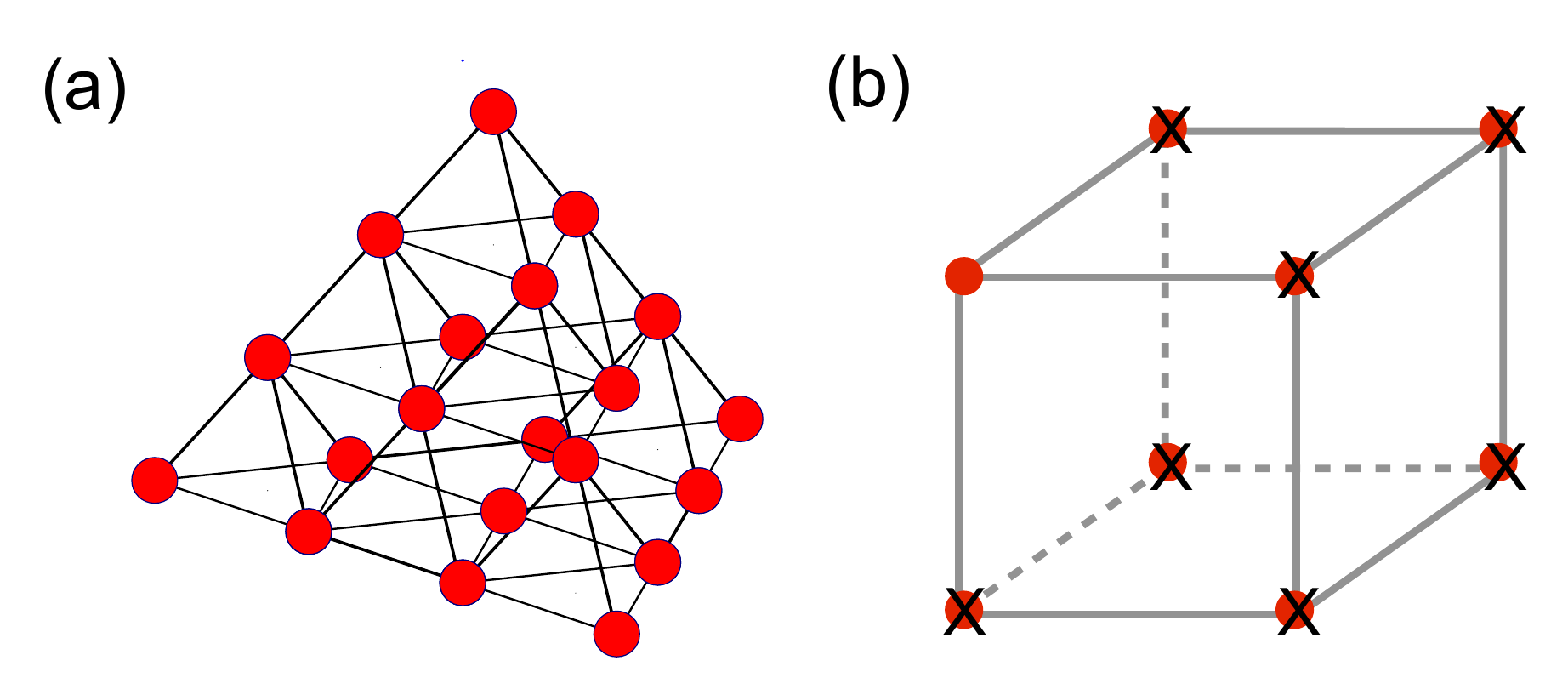}
\caption{(Color online) (a) The tetrahedral lattice. (b) A real-space RG on a cubic lattice.
} 
\label{fig_cube}
\end{figure}

One may naturally hope to develop three-dimensional real-space RG by considering a tetrahedral lattice with blocks of four spins (Fig.~\ref{fig_cube}(a)). However, there seems to be no simple scale-invariant covering of the tetrahedral lattice in general. Therefore, one needs to break translation symmetry or isotropy of the lattice for successful renormalization, which might lead to imprecise estimates of $\nu$. One might also think of the Pyrochlore lattice, but the number of nearest neighbors changes after renormalization. Finally, one may consider a renormalization of a cubic lattice as depicted in Fig.~\ref{fig_cube}(b). By using the geometric mean of coupling strengths, we obtain
\begin{align}
\nu =0.4519
\end{align}
which is not far from the value $\nu=1/2$ predicted by the mean-field theory. However, the estimate is not as successful as in two dimensions.

\begin{figure}[htb!]
\centering
\includegraphics[width=0.90\linewidth]{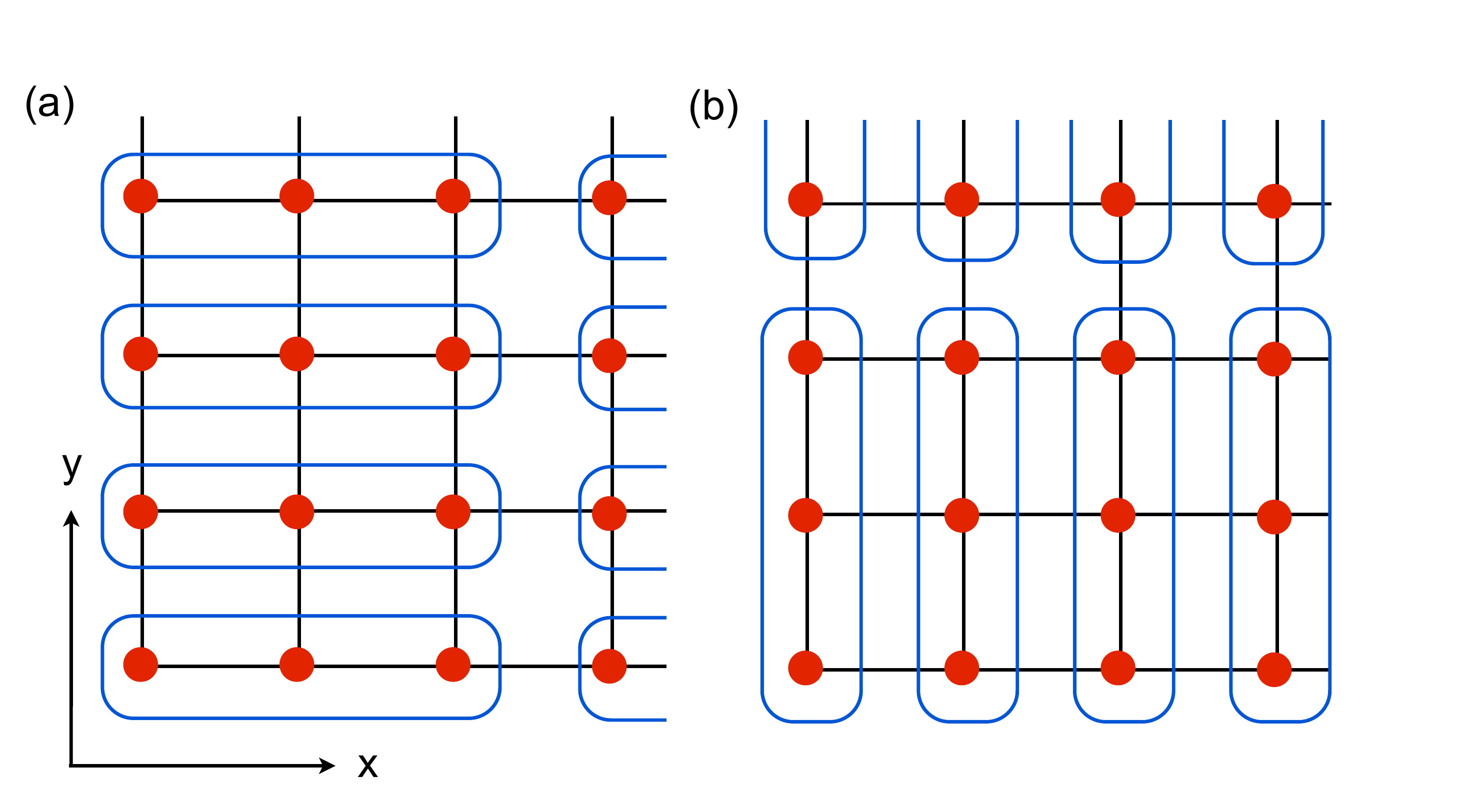}
\caption{(Color online) A hybridized RG scheme of the quantum Ising model on a square lattice. (a) Renormalization in the $\hat{x}$ direction. (b) Renormalization in the $\hat{y}$ direction.
} 
\label{fig_2Da}
\end{figure}

Here, we consider a hybrid of lower-dimensional renormalization procedures to form effectively isotropic three-dimensional renormalization scheme, based on an idea by Miyazaki \textit{et al}~\cite{Miyazaki11}. In order to gain some insight into the method, let us begin with a hybridized RG scheme on a two-dimensional square lattice (Fig.~\ref{fig_2Da}). The RG scheme is an iteration of the following two steps:
\begin{enumerate}
\item Renormalize the spins in the $\hat{x}$ direction by grouping them in blocks of $b$ spins. 
\item Renormalize the spins in the $\hat{y}$ direction by grouping them in blocks of $b$ spins.
\end{enumerate}
Here we choose $H_{in}= -J\sum_{j=1}^{b-1}Z_{j}Z_{j+1}-h\sum_{j=1}^{b-1}X_{j}$ by removing one on-site term $X$ from the block of $b$ spins. After one iteration of steps 1-2, the system is rescaled by a factor of $b$. We then solve an RG equation for the geometric mean of coupling strengths. An estimate for the two-dimensional quantum Ising model is
\begin{align}
\nu=0.6211 \quad  (b=2) \quad \nu = 0.6315 \quad (b=3)
\end{align}
which is in good agreement with numerical estimates. It is worth emphasizing that one obtains better estimate by grouping three spins instead of only two. 

Next, consider the following RG transformation on a three-dimensional cubic lattice. 
\begin{enumerate}
\item Renormalize the spins in the $\hat{x}$ direction by grouping them in blocks of $b$ spins. 
\item Renormalize the spins in the $\hat{y}$ direction by grouping them in blocks of $b$ spins.
\item Renormalize the spins in the $\hat{z}$ direction by grouping them in blocks of $b$ spins. 
\end{enumerate}
By solving an RG equation for the geometric mean of coupling strengths, we obtain
\begin{align}
\nu=0.5014 \quad  (b=2) \quad \nu = 0.5138 \quad (b=3)
\end{align}
which is consistent with $\nu=1/2$ predicted by the mean-field theory. In three dimension, it is better to group spins in blocks of two. The precision of estimating $\nu$ in two and three dimensions is comparable.

In order to further increase the precision of estimates, we consider a hybrid of two-dimensional and one-dimensional renormalization procedures. Consider layers of triangular lattices piled on top of each other, filling the three-dimensional space, as in Fig.~\ref{fig_3D}(a). We consider the following RG transformation:

\begin{figure}[htb!]
\centering
\includegraphics[width=0.95\linewidth]{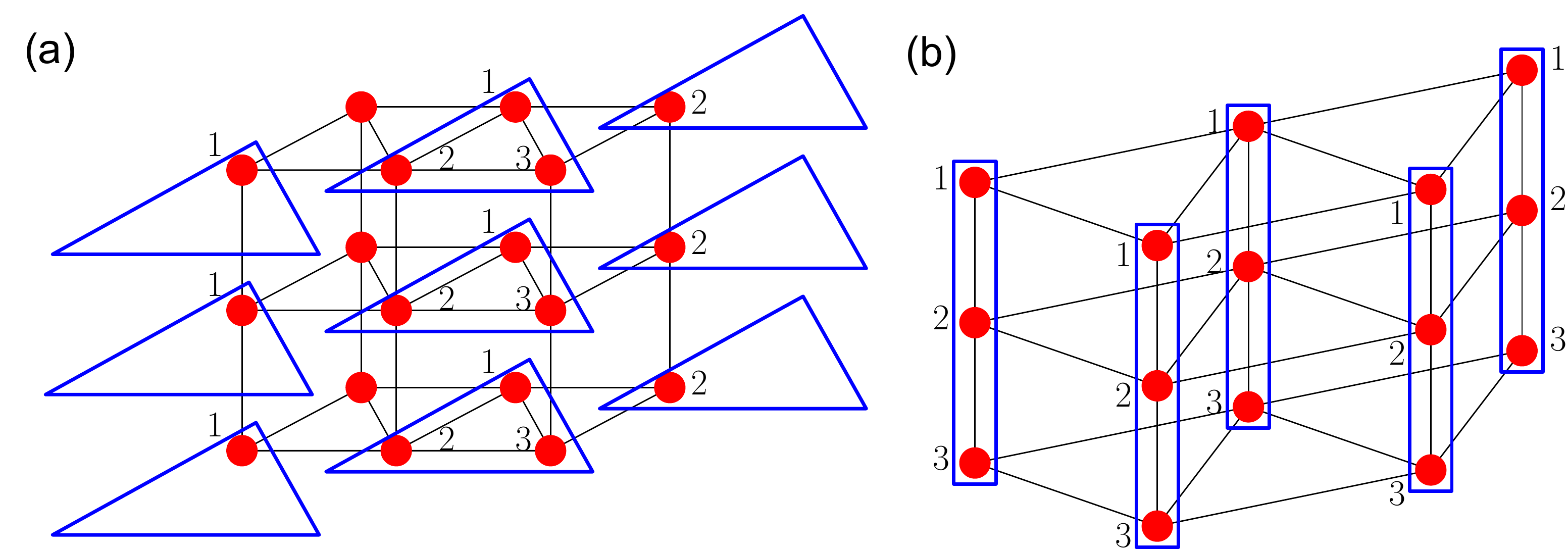}
\caption{(Color online) Renormalization in three dimension. (a) Division of triangular lattices into blocks of three spins. (b) Grouping spins into vertical blocks of three spins. 
} 
\label{fig_3D}
\end{figure}

\begin{enumerate}
\item Within each horizontal layer of triangular lattices, renormalize the Hamiltonian via the two-dimensional canonical method. 
\item Renormalize the spins in the vertical direction by grouping them in blocks of three spins. 
\item Repeat the step 1.
\end{enumerate}

Note that we perform two-dimensional renormalization twice since a single renormalization rescales the system only by a factor of $\sqrt{3}$ in a plane. After one iteration of steps 1--3, the system is rescaled by a factor of three as a whole where $27$ spins are grouped in total. An estimate for the three-dimensional quantum Ising model is
\begin{align}
\nu = 0.4986
\end{align}
which is close to $\nu=1/2$ predicted by the mean-field theory. 

Finally, we apply the same RG scheme to the quantum Potts model:
\begin{align}
\nu =0.4337.
\end{align}
We were not able to find any previous work which numerically estimates this quantity. 

\section{Real-space RG on fractal lattices}

\begin{figure}[htb!]
\centering
\includegraphics[width=0.95\linewidth]{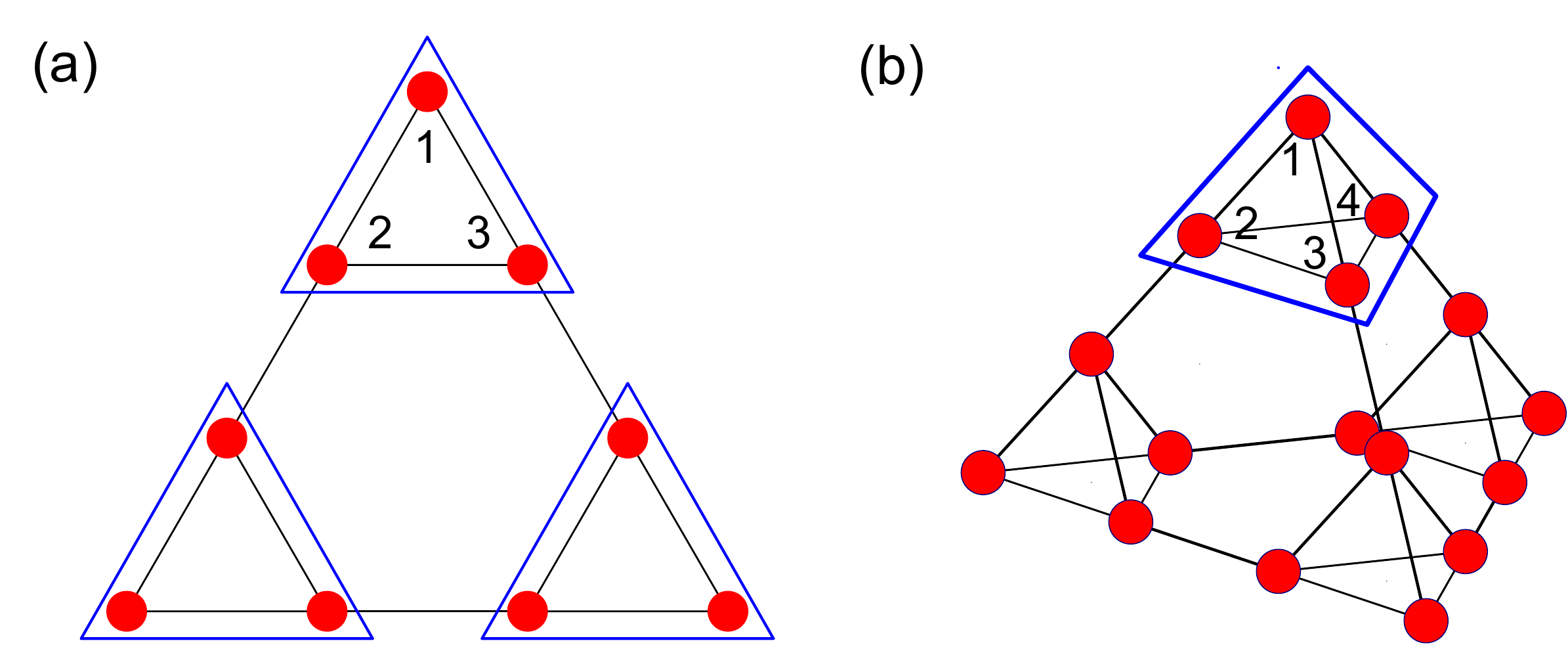}
\caption{(Color online) (a) The Sierpi\'{n}ski triangle in $\mathbb{R}_{2}$. (b) The Sierpi\'{n}ski pyramid in $\mathbb{R}_{3}$. 
} 
\label{fig_fractal}
\end{figure}

Finally, we consider the quantum Ising and Potts models on fractal lattices. To be specific, we consider fractal lattices based on generalized Sierpi\'{n}ski pyramid in $m$ spatial dimensions, whose Hausdorff dimension is $\log (m+1)/ \log 2$. Note that, for $m=2$ and $m=3$, the lattices resemble the well-known Sierpi\'{n}ski triangle and pyramid, respectively, as depicted in Fig.~\ref{fig_fractal}. It has been numerically demonstrated that the quantum Ising model on Sierpi\'{n}ski pyramid undergoes second-order quantum phase transition where scaling relations are satisfied with good precisions for $m=2,3$~\cite{Beni14a}. Here, we estimate the correlation length exponent $\nu$ through real-space RG schemes and analyze its dependence on the spatial dimension. 
  
We group spins into blocks of $m+1$ spins which form $m$-simplices as shown in Fig.~\ref{fig_fractal}. We choose 
\ba{
H_{in}=-\sum_{j=2}^{m+1}J_{1i}Z_{1} Z_{j}-h \sum_{j=2}^{m+1} X_{j}.
}
Note that $H_{in}$ is doubly degenerate due to the following symmetry operators
\ba{
\bar{X}=X_{1}\qquad \bar{Z}=Z_{1}\cdots Z_{m+1}
}
and does not have any loops of ferromagnetic terms. Then we obtain
\begin{subequations}
\label{e:jtransf}
\begin{align}
k'_{1i} & = k^2_{1i}\prod_{j\neq 1,i}\sqrt{1+k^2_{1j}}, \\
k'_{ij} & = k_{ij} k_{1i} k_{1j} \prod_{l\neq 1,i,j} \sqrt{1+k^2_{1l}} .
\end{align}
\end{subequations}
For the fixed point, we require
\begin{equation}
\label{e:jcondition}
\prod_{i\neq j}k_{ij}=\prod_{i\neq j}k'_{ij}. 
\end{equation}
Since we have started with an isotropic lattice, $k_{ij}=k$, and we set $k'$ to be the geometric mean of  $k'_{ij}$'s, i.e.  $k'\equiv\left(\prod_{i<j} k'_{ij}\right)^{2/m(m+1)}$. Then, from Eq.~(\ref{e:jtransf}) we obtain
\begin{align}
k &\rightarrow k' = k^{(3m+1)/(m+1)} (1+k^2)^{m(m-1)/2(m+1)}.
\end{align}
By solving for the fixed point, we obtain estimates of $\nu$, as summarized in Table~\ref{t:sierpinski}. In~\cite{Beni14a}, the quantum Ising model on the Sierpi\'{n}ski pyramid for $m=2,3$ has been analyzed via Monte-Carlo simulations, which yield $\nu=0.76, 0.66$ for $m=2,3$ respectively. 

\begin{figure}[htb!]
\centering
\includegraphics[width=0.95\linewidth]{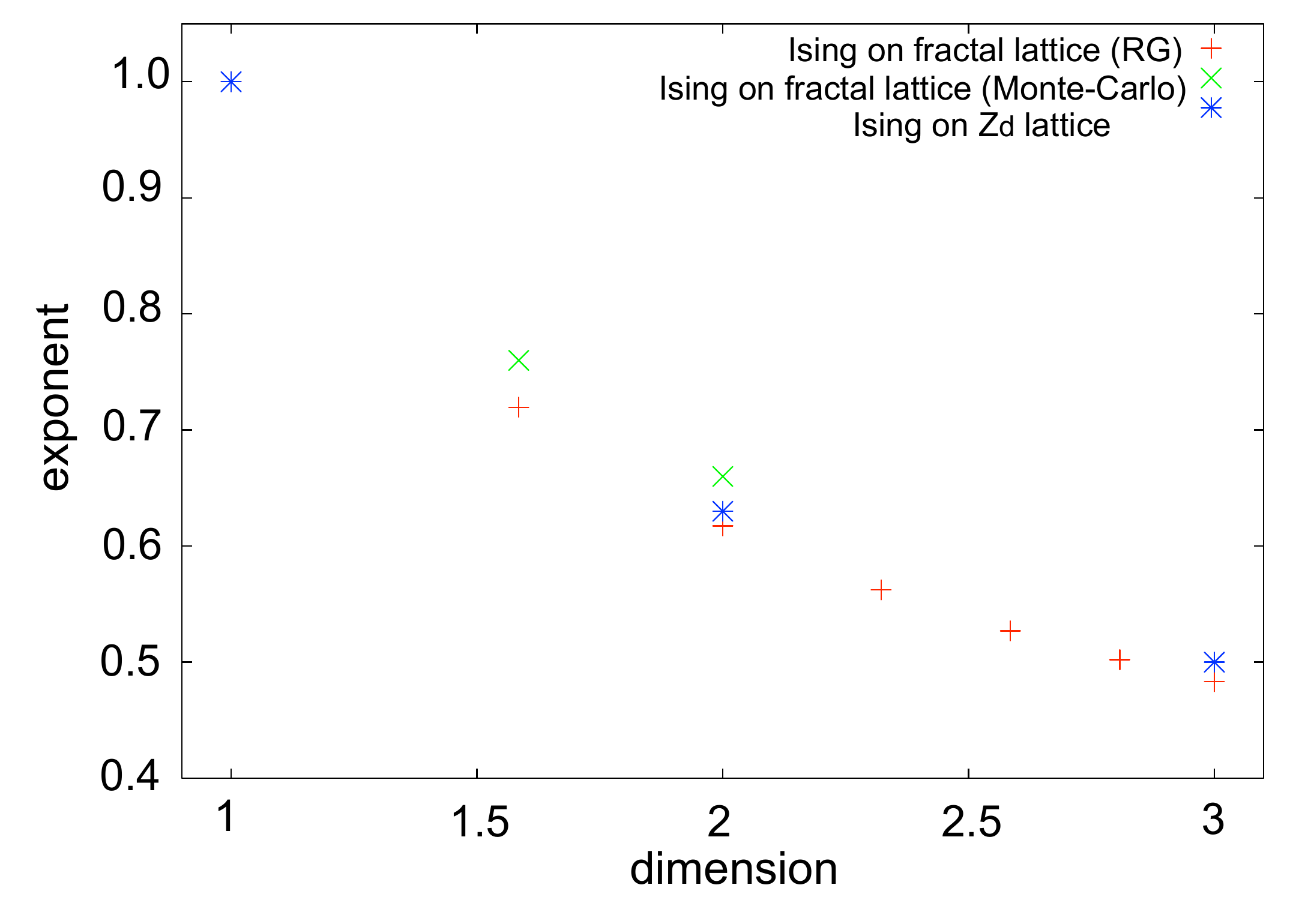}
\caption{(Color online) Correlation length exponent and Hausdorff dimension of fractals.
} 
\label{fig_plot}
\end{figure}

As seen in Fig.~\ref{fig_plot}, values of $\nu$ in fractal lattice models interpolate those of integer-dimensional models. This implies that $\nu$ is mostly determined by symmetry and spatial dimension of the system. Data points for the Ising model in Fig.~\ref{fig_plot} can be fitted well with a function $\nu\propto 1/(d+1)$. This observation together with the Widom scaling $2-\alpha=\nu (d+1)$ suggests that the specific heat critical exponent $\alpha$ does not change significantly with dimension of the analyzed lattices. For the Ising model, it can be approximated by $\alpha\approx 0.12$.

\begin{table}[h]
\centering
\begin{tabular}{c|c|c|c|c}
 \vspace*{-1.5mm} 
 space & Hausdorff dimension & classical & $\nu_{Ising}$ & $\nu_{Potts}$ \\
& of fractal& dimension & &  \\
\hline
$\mathbb{R}^2$ & $\log 3/\log 2$ & 2.5850 & 0.7196 & 0.6213\\
$\mathbb{R}^3$ & $\log 4/\log 2$ & 3 & 0.6174 & 0.5390 \\
$\mathbb{R}^4$ & $\log 5/\log 2$ & 3.3219 & 0.5623 & 0.4946 \\
$\mathbb{R}^5$ & $\log 6/\log 2$ & 3.585 & 0.5270 & 0.4662 \\
$\mathbb{R}^6$ & $\log 7/\log 2$ & 3.8074 & 0.5021 & 0.4462 \\
$\mathbb{R}^7$ & $\log 8/\log 2$ & 4 & 0.4832 & 0.4311\\
\end{tabular}
\caption{\label{t:sierpinski} The correlation length critical exponent $\nu$ for the quantum Ising and Potts models on the generalized Sierpi\'{n}ski pyramids.}
\end{table}

\section{Summary}

In this paper, we have demonstrated that real-space RG schemes can predict values of the correlation length exponent $\nu$ with amazingly good precision which is comparable to estimates from more intricate methods which have been traditionally used. Given tremendous amount of time, resource and knowledge that have been accumulated and devoted to studies of quantum critical phenomena over decades, the success of our RG scheme, which is fairly simple and analytical, seem quite surprising. An analytical solution of ($2+1$)-dimensional Ising model is one of the greatest open questions in physics. Our result may serve as a small, but an important step in finding a method for solving analytically or approximating the ($2+1$)-dimensional Ising model. We were not able to fully justify the success behind our calculations. Yet, we have identified several crucial conditions under which RG schemes work successfully via throughout investigation of various real-space RG schemes. Taking these conditions into consideration, our RG scheme seems among a few reasonable choices. It may be an interesting future problem to further generalize real-space RG schemes to various models of quantum many-body systems. 

An important question we did not address in this paper concerns the way of computing the magnetic exponents and transition points through real-space RG schemes. Real-space RG schemes developed in this paper fail to determine the exact value of the magnetic exponent. One possible modification is to investigate how the energy scale changes under renormalization. In one-dimensional real-space RG scheme considered in Eq.~(\ref{eq:1dim}), the fixed point solution corresponds to $J'=J/\sqrt{2}$ and $h'=h/\sqrt{2}$ with $J=h$. This implies that the energy gap between the ground state and the first excited state scales as $O(1/\sqrt{L})$ in the thermodynamic limit. Yet, a correct one-dimensional quantum critical system would have an energy gap which scales as $O(1/L)$. By developing a real-space RG scheme which correctly reproduces the scaling of energy gap, one may be able to obtain better estimates of some critical exponents. As for transition points, we note that the fixed points obtained in our RG schemes are different from actual quantum phase transition points in general. We believe that this does not invalidate our RG schemes since critical exponents are determined by long-range properties only and do not depend on details of the system. Indeed, in $\phi^4$-theory approach, the effective description captures long-range properties of ferromagnets correctly while it does not give precise estimation of transition points.

\section*{Acknowledgment}

We thank Glen Evenbly for helpful discussion and comments. BY is supported by the David and Ellen Lee Postdoctoral fellowship. We acknowledge funding provided by the Institute for Quantum Information and Matter, an NSF Physics Frontiers Center with support of the Gordon and Betty Moore Foundation (Grants No. PHY-0803371 and PHY-1125565).


\end{document}